\documentclass[12pt]{article}
\usepackage{amsmath, amsthm, amssymb}

\title{\bf Cooper Pair Boxes Weakly Coupled to External Environments}

\author{Fabio Benatti$^{a,b}$, 
Roberto Floreanini$^{b}$, John Realpe-G\'omez$^c$\\
\small ${}^a$Dipartimento di Fisica Teorica, Universit\`a di Trieste,
34014 Trieste, Italy\\
\small ${}^b$Istituto Nazionale di Fisica Nucleare, Sezione di Trieste,
34014 Trieste, Italy\\
\small ${}^c$The Abdus Salam International Center for Theoretical Physics,
34014 Trieste, Italy
}

\date{\null}

\begin{document}

\maketitle

\begin{abstract}
\noindent
We study the behaviour of charge oscillations in
Superconducting Cooper Pair Boxes
weakly interacting with an environment.
We found that, due to the noise and dissipation induced by the environment,
the stability properties of these nanodevices differ according to whether the
charge oscillations are interpreted
as an effect of macroscopic quantum coherence, or semiclassically
in terms of the Gross-Pitaevskii equation. 
More specifically, occupation number states, used in the
quantum interpretation of the oscillations, are found to be much more
unstable than coherent ones, typical of the semiclassical explanation.
\end{abstract}

\section{Introduction}

Low-capacitance Josephson-junction devices have recently attracted a wide interest,
both theoretically and experimentally, particularly in view of the possibility of identifying
macroscopic quantum phenomena in their behaviour.
In this respect, one of the circuits that have gained great attention is the 
so-called Superconducting Cooper Pair Box (SCB), with an increasing number of experiments aimed at supporting 
a qubit interpretation of its evolution ({\it e.g.}, see \cite{Naka1}-\cite{Steffen}).

The SCB is a circuit consisting of two superconducting electrodes 
linked through a Josephson junction and coupled, capacitively, to a voltage source. 
One of the superconducting electrodes is assumed to be small enough 
for the charging energy to play the main role. In this situation, there is the possibility 
of tunnelling electrons one by one through the Josephson junction, allowing 
for external control of charge oscillations \cite{Naka1}-\cite{Grabert}. 
Because of the large number of Cooper pairs
in the two electrodes, these oscillations have been interpreted as genuine (macroscopic) 
manifestation of quantum coherence. Nevertheless, the experimental data allow for an
equivalent explanation of the oscillations in terms of a semiclassical behavior of the system. 
Two different theoretical scenarios can then be used in modelling the observed charge oscillations:
one is the so-called {\it quantum phase model} \cite{Makhlin,Wendin} that essentially 
describes a quantized non-linear harmonic oscillator, while the other is a {\it mean-field model} 
that leads to a Gross-Pitaevskii like equation \cite{Alicki1,Alicki2}; the quantum phase model describes
the system in terms of occupation number (Fock) states, on the other hand, in the mean-field formulations
coherent-like states naturally appear.

More in general, the difficulties for a distinction between classical and quantum behavior 
in Josephson-junction based devices have been pointed out before in 
\cite{Gronbech1}-\cite{Gronbech3}. 
There, it is shown that classical nonlinear-oscillator-like 
models can reproduce some experimental results that have 
been previously attributed to genuine quantum macroscopic behavior. 

In the present work, we limit our consideration to the SCB charge oscillations and 
address the problem of discriminating between their quantum and classical 
behavior from an open quantum systems perspective, {\it i.e.} when the SCB is
immersed in a weakly coupled external environment. In this case, the SCB dynamics
is no longer unitary; instead, it is described by a generalized evolution of semigroup
type (a so-called {\it quantum dynamical semigroup}), that incorporates effects of
dissipation and decoherence induced by the environment~\hbox{\cite{K}-\cite{BP}}. 
More specifically, we will focus our attention
on the stability properties of the system against the induced environmental noise.
A preliminary investigation on this issue has been reported in \cite{ABF}, using the
so-called {\it singular coupling limit} \cite{Gorini}, corresponding to a specific environment,
with stochastic or white noise correlations.
Here instead, we shall consider a generic environment and study the
SCB stability properties in the {\it weak coupling limit} \cite{Davies}, a procedure that
implements in a physically consistent way the weak interaction of the SCB with the environment.

The main point of our investigation is that any coupling with an external environment
needs a microspic description: only in this case one can derive a master equation valid in any
physical situation \cite{AL}. Once a general master equation is obtained, 
the choice between the two possible explanations
of the observed SCB charge oscillations depends on the stability against noise of the
states on which the two different models are based. In this respect, our approach
differs for instance from the one in \cite{Choi}, which assumes the validity
of the {\it quantum phase model} and describes the interaction with the environment
by means of a spin-boson model.

We found that although both quantum phase and mean-field models predict decoherence, 
they give rise to quite different decay properties: occupation number states
(used in the former model) turn out to be much less stable than coherent ones
(typical of the latter), thus confirming
the results of \cite{ABF}. Therefore, the measure of the decay rate of the 
SCB charge oscillations in presence of noise
would in principle allow to discriminate between the two models.

\section{SCB oscillations: two approaches}

In a suitable regime and in absence of the environment, 
the dynamics of a SCB can be effectively modelled by a Bose-Hubbard Hamiltonian \cite{Cirac,Lewe}
(henceforth, the small electrode will be labeled by $L$, while 
the other much bigger one by $R$); in terms of bosonic creation and annihilation operators
$\hat{a}_i^\dag$, $\hat{a}_i$, $i=L,R$ in the two electrodes, one can write
\begin{equation}
H_0=E_C\, \big(\hat{a}_L^\dag \hat{a}_L\big)^2 + U_L\, \hat{a}_L^\dag \hat{a}_L 
+ U_R\,  \hat{a}_R^\dag \hat{a}_R - K\, \big(\hat{a}_L^\dag \hat{a}_R+\hat{a}_L \hat{a}_R^\dag \big)\ ,
\label{h0}
\end{equation}
where the quadratic term $E_C \big(\hat{a}_L^\dag \hat{a}_L\big)^2$ 
accounts for Coulomb repulsion in the small island 
(the one in the much larger electrode $R$ can be neglected), 
$U_i\hat{a}_i^\dag \hat{a}_i$, $i=L,R,$ are potential contributions, 
while the last one is the tunneling term. 

Due to the conservation of charge in the SCB, the Hamiltonian \eqref{h0} must be restricted to the subspace with a constant total number of particles $N=n_L+n_R$. Two effective descriptions
of the dynamics of the SCB can then be obtained, the quantum phase model and the mean-field one.

In the former case, the relevant states are occupation number (Fock) 
states with a constant total number of particles
\begin{equation}
|n\rangle\equiv |n_L=n,n_R=N-n\rangle = 
\frac{1}{\sqrt{n!(N-n)!}}\big(\hat{a}_L^\dag\big)^n
\big(\hat{a}_R^\dag\big)^{N-n}\, |\mathrm{vac}\rangle\ ,
\label{fock}
\end{equation}
where $|\mathrm{vac}\rangle$ is the vacuum state. Ignoring a constant term, 
the Hamiltonian in \eqref{h0} can be rewritten as
\begin{equation}
H_0=E_C \left(\hat{n}_L -\bar{n}-n_g\right)^2
- K(\hat{a}_L^\dag \hat{a}_R+\hat{a}_L \hat{a}_R^\dag )\ ,
\label{h0-1}
\end{equation}
where $\hat{n}_L=\hat{a}_L^\dagger \hat{a}_L$ is the number operator in electrode $L$, 
while $\bar{n}$ is the corresponding
average number; in typical experimental situations,
one has $\bar n\approx 10^8$ \cite{Naka1}-\cite{Steffen}.
Thus, the difference
$\hat{n}'\equiv\hat{n}_L -\bar{n}$ measures the excess of Cooper pairs (hence of charge) in
the same electrode. 
The parameter
$n_g=(U_R-U_L)/2E_C - \bar{n}$ is connected with the potential energy difference
across the Josephson junction, or equivalently with the average charge in the gate
capacitor, and it can be controlled externally through the voltage source.

In the occupation number representation,
the Hamiltonian \eqref{h0-1} takes the form
\begin{equation}
H_0=E_C\sum_{n=0}^N\left(n-\bar{n}-n_g\right)^2\vert n\rangle\langle
n\vert -\,E_J\,\sum_{n=0}^N\Bigl(\vert n\rangle\langle n+1\vert\,+\,\vert
n+1\rangle\langle n\vert\Bigr)\ ,
\label{h0-2}
\end{equation}
where we have set $K\sqrt{(n+1)(N-n)}\simeq K\sqrt{\overline{n}(N-\overline{n})}
\equiv E_J$
since we are interested in low lying states, for which $|n -\overline{n}|\equiv |n'|\ll \overline{n}$.
Introducing the conjugated operators $\hat{n}'$ and $\hat{\varphi}$, obeying
the canonical commutation relations, $[\hat{\varphi},\ \hat{n}']=i$, 
so that $\mathrm{e}^{\pm i\hat{\varphi}}$ 
decrease (increase) $n'$ by 1, 
the expression in \eqref{h0-2} is equivalent to the so-called
quantum phase model Hamiltonian \cite{Makhlin,Wendin}
\begin{equation}
H_0=E_C\left(\hat{n}'-n_g\right)^2- E_J\, \cos\hat{\varphi}\ .
\label{h0-3}
\end{equation}

By adjusting the gate voltage so that $n_g\approx 1/2$, 
one enters a particular situation (resonance) where only the occupation number states 
$|n\rangle=|n'+\bar n\rangle$ for which $n'=\,0$ and $n'=1$ play a role 
and are strongly coupled by the Josephson junction. 
In this case, 
the Hamiltonian \eqref{h0-2} can then be approximated by a two level Hamiltonian
\begin{equation}
\hat{H}_{eff}=-\frac{1}{2}[E_C(1-2n_g)\sigma_Z +E_J\sigma_X]\ ,
\end{equation}
where $\sigma_X$ and $\sigma_Z$ are Pauli matrices. The effective two level system will thus display coherent oscillation with frequencies given by
\begin{equation}
\omega_q=\Delta E=\sqrt{E_C^2(1-2n_g)^2+E_J^2}\approx E_J\ ,
\label{q-frequency}
\end{equation}
where the expression in the middle has been approximated to $E_J$ because we are working close to
resonance. 

A different description of the system can be given by treating the starting microscopic
Hamiltonian \eqref{h0} in a mean-field approach. This approximation is justified by the large number 
of Cooper pairs in the two electrodes, all in the same condensed state.
This situation can be properly described by the product of $N$ single Cooper pair states
\begin{equation}
|\Psi\rangle_N = \frac{1}{\sqrt{N!}}
\left(\psi_L\hat{a}_L^\dag+\psi_R \;\hat{a}_R^\dag\right)^N|\mathrm{vac}\rangle =\sum_{n=0}^N
C_n|n\rangle\ , 
\label{psiN}
\end{equation}
with
\begin{equation}
C_n=\sqrt{\frac{N!}{n!(N-n)!}}\;\psi_L^{n}\,\psi_R^{N-n},
\label{Cn}
\end{equation}
where $\psi_L$ and $\psi_R$ can be interpreted as the condensed wave function in each side of the junction (with $|\psi_L|^2 + |\psi_R|^2=1$). 
The dynamics of $\vert\Psi\rangle_N$ follows the standard
Schr\"odinger equation governed by the Hamiltonian \eqref{h0}
\begin{equation}
\label{Schr-eq}
i\frac{{\rm d}}{{\rm d}t}\vert\Psi(t)\rangle_N=H_0
\vert\Psi(t)\rangle_N\ .
\end{equation}
The two sides of this equation can be explicitly computed with the help of the following relations
\begin{subequations}
\begin{align}
&\hat{a}_L|\Psi(t)\rangle_N =\sqrt{N}\psi_L|\Psi(t)\rangle_{N-1},\\
&\hat{a}_R|\Psi(t)\rangle_N =\sqrt{N}\psi_R|\Psi(t)\rangle_{N-1},\\
\label{psiL}
&n_L \equiv {_N\langle} \Psi(t)|\hat{a}_L^\dag \hat{a}_L|\Psi(t)\rangle_{N}=N|\psi_L|^2,\\
\label{psiR}
&n_R \equiv N-n_L={_N\langle} \Psi(t)|\hat{a}_R^\dag \hat{a}_R|\Psi(t)\rangle_{N}=N|\psi_R|^2\ .
\end{align}
\end{subequations}
By further using 
$\hat{a}_L^\dag\hat{a}_L\approx\langle\hat{a}_L^\dag\hat{a}_L \rangle=n_L=N\, |\psi_L(t)|^2$,
justified by the mean-field approximation, one finds that the
product state $\vert\Psi(t)\rangle_N$ is a solution of the evolution equation 
(\ref{Schr-eq})
if the amplitudes $\psi_i(t)$ solve the Gross-Pitaevskii equations for
the two component order parameter $(\psi_L,\psi_R)$:
\begin{align}
i\dot{\psi}_L(t) & =[U_L+NE_C|\psi_L(t)|^2]\psi_L(t)-K\psi_R(t)\ ,\\
i\dot{\psi}_R(t) & =U_R\psi_L(t)-K\psi_L(t)\ .
\end{align}
By setting $\psi_i=\sqrt{n_i/N}\mathrm{e}^{i\theta_i}$ 
and using the conservation of the total number of particles $n_L+n_R=N$, 
the equations above can be written in terms of the Hamiltonian function (up to an additive constant)
\begin{equation}
\mathcal{H}(\theta , n_L)={E_C}(n_L-\bar{n}-{n}_g)^2-{E}_J\cos\theta
\end{equation}
where the same definitions as before for $n_g$ and $E_J$ have been used, together with
$\theta=\theta_L-\theta_R$. This Hamiltonian function
describes semiclassical charge oscillations with frequency 
\begin{equation}\label{c-frequency}
\omega_c =\sqrt{2E_C{E}_J}\ .
\end{equation}
Notice that the state $|\Psi\rangle_N$ in \eqref{psiN} behaves like a coherent state; indeed, 
due to the large numbers involved $N\gg \bar{n}\gg 1$, 
we can replace the coefficients $C_n$ of $|\Psi\rangle_N$ by a Poisson distribution
so that in the limit of $N$ large, one can write:
\begin{equation}
|\Psi\rangle_N\approx |\alpha\rangle\equiv\sum_{n=0}^\infty \left[\frac{\bar{n}^n}{n!}\,
\mathrm{e}^{-\bar{n}}\right]^{\frac{1}{2}}\mathrm{e}^{-in\theta}| n\rangle\ ,
\label{coherent}
\end{equation}
which is indeed a coherent state satisfying
$$\hat{b}|\alpha\rangle =\alpha|\alpha\rangle, \hspace{1 cm} \alpha =\sqrt{\bar{n}}\;\mathrm{e}^{-i\theta}$$
where $\hat{b}$ is the annihilation operator for a fictitious nonlinear oscillator with eigenvectors 
$| n\rangle $, 
$$\hat{a}_L^\dag\hat{a}_R| n\rangle =\sqrt{N-n}\;\hat{b}^\dag | n\rangle ,\hspace{0.5cm}\hat{a}_L\hat{a}_R^\dag| n\rangle =\sqrt{N-n+1}\;\hat{b}| n\rangle . 
$$
Hence, the oscillations observed in a SCB might be the result of a semiclassical behaviour 
described by the coherent state $|\Psi\rangle_N\approx|\alpha\rangle$, rather than a manifestation of quantum coherence at the macroscopic level.

\section{SCB with noise: weak coupling limit}

From the previous discussion, it follows that there are two possible scenarios to describe charge oscillations in SCB, namely the quantum phase model, which is based on a purely quantum description, and the mean-field one which is semiclassical in nature. They both are qualitatively consistent with the experimental data: 
if these could be used to measure the oscillation frequency, a direct discrimination between the
two approaches would be possible; however, the present experimental accuracy does not seem
to allow this. Instead, we shall show that the two approaches give different decoherence patterns when the SCB is weakly coupled to an environment which acts as a source of noise and dissipation.%
\footnote{Within the qubit interpretation, different aspects of decoherence
phenomena in SCB behaviour have been discussed in \cite{Choi}.}
In \cite{ABF} a specific model of environment has been considered, which
required the so called {\it singular coupling limit} technique;
in the following we shall study the effects of a more general source of dissipation,
using the {\it weak coupling limit} approach.

We describe the weak coupling of the SCB to an external environment
by means of the total Hamiltonian
\begin{equation}
\label{h}
H=H_0 + H_E +\lambda\Bigl(
a_1a_2^\dagger \otimes B\,+\,a^\dagger_1a_2 \otimes B^\dagger\Bigr)\ ,
\end{equation}
where $H_0$ is the microscopic Bose-Hubbard Hamiltonian \eqref{h0}, $H_E$ is the Hamiltonian of the environment, 
$\lambda\ll1$ is a small coupling
constant and $B$ is a suitable environment operator. 
We shall assume the environment to be in an equilibrium
state $\rho_E$,  with two-point correlation
functions, $\langle B^\dagger(t)\, B\rangle_E\equiv {\rm Tr}_E[\rho_E\, B^\dagger(t)\, B]$ 
and similar ones, that decay fast enough (for details, see \cite{Davies}).
A heat bath, with $\rho_E\simeq e^{-\beta H_E}$, is a typical 
example of environment fulfilling these conditions: specifically, it can be identified with
the cloud of non-condensed electrons in the two SCB electrodes.

This very general situation provides the setting for the so-called 
{\it weak coupling limit} \cite{Davies}, a physically consistent
and mathematically precise procedure leading to an evolution equation for the SCB density matrix
$\rho$ in Kossakowski-Lindblad form \cite{K}-\cite{BP}:
\begin{equation}
\label{KL}
\frac{\partial \rho}{\partial t}=-i[H_0+H^{(2)},\rho]+\mathcal{D}[\rho]\ .
\end{equation}
The contribution $H^{(2)}$ is an environment induced Hamiltonian correction to the starting
system Hamiltonian $H_0$, whose explicit expression will not be relevant in the sequel;
on the other hand, the term $\mathcal{D}$ describes the non-Hamiltonian effects of the environment on the dynamics of the SCB, which typically result in dissipation and noise. 
In deriving \eqref{KL}, we have assumed to work in the experimentally relevant 
regime in which the effective Josephson coupling $E_J$ is of the same
order of magnitude of the dissipative effects, that start to become relevant at order $\lambda^2$.
In this regime, we explicitly find:
\begin{equation}
\label{D}
\begin{split}
\mathcal{D}[\rho] = \lambda^2\sum_{n=0}^N&\left\{ h(\omega_n)\left(W^\dag(n)\, \rho\,  W(n)
-\frac{1}{2}\Big\{W(n)W^\dag(n), \, \rho\Big\}\right) +\right.\\ 
& \left. +\kappa(\omega_n)\left(W(n)\, \rho\,  W^\dagger(n) 
-\frac{1}{2}\Big\{W^\dagger(n)W(n),\,\rho\Big\}\right)\right\}\ ,
\end{split}
\end{equation}
where 
\begin{equation}
W(n)=\sqrt{(n+1)(N-n)}\, |n\rangle\langle n+1|\ ,
\end{equation}
and 
\begin{equation}
\label{Fourier}
h(\omega_n)=\int_{-\infty}^\infty\mathrm{d}t\, \mathrm{e}^{-it\omega_n}\, 
\langle B(t)\, B^\dagger\rangle_E\ ,
\qquad 
\kappa(\omega_n)=\int_{-\infty}^\infty\mathrm{d}t\, \mathrm{e}^{it\omega_n}\, 
\langle B^\dagger(t)\, B\rangle_E\ ,
\end{equation}
are the Fourier transform of the environment correlations with respect to the frequencies
\begin{equation}
\label{omega-n}
\omega_n =E_C[2(n-\bar{n}-n_g)+1]\ ,
\end{equation}
where $n$ is, as before, the actual number of Cooper pairs in the island $L$.%
\footnote{There is an additional contribution to $\mathcal{D}$ that arises strictly at resonance, 
{\it i.e.} when $n_g=1/2$. For simplicity, we have omitted it, since it will play no role in the following discussions; indeed, its contribution to the decay rates is suppressed by a factor 
$1/\sqrt{\bar{n}}\sim 10^{-4}$ with respect to the dominant one coming from \eqref{D} (for further details,
see \cite{John}).}

The stability properties of any initial SCB pure state,
$\rho=|\psi\rangle\langle\psi |$, can be studied using the master equation \eqref{KL}.
Indeed, first note that the difference between two quantum states can be estimated by looking 
at the part of one of them that is orthogonal to the other. 
Thus, we can quantify the degree of stability of any state $\rho$ by 
measuring how fast it starts deviating from itself, {\it i.e.} by
evaluating the initial rate of variation $\Gamma$ of the orthogonal contribution to itself.
Explicitly,
\begin{equation}
\label{Gamma}
\Gamma\equiv\left.\frac{\mathrm{d}}{\mathrm{d}t}\mathrm{Tr}\left[\left(\boldsymbol{1}-|\psi\rangle\langle\psi |\right)\rho(t)\right]\right|_{t=0}=-\langle\psi |\frac{\partial \rho(t)}{\partial t}|\psi\rangle\bigg|_{t=0}
=-\langle\psi |\mathcal{D}\big[|\psi\rangle\langle\psi |\big]|\psi\rangle\ ,
\end{equation}
where in the last equality we have used \eqref{KL}, 
taking into account that the Hamiltonian contribution vanishes.
Roughly speaking, a generic state is expected to decay exponentially with 
a rate given by $\Gamma$, so that the bigger $\Gamma$ is, the faster it decays, the less stable 
it is. In the specific case under study, we expect to find an appreciable difference between the decay 
rates of occupation number (Fock) and coherent states that might shed light on the suitability of 
the two discussed models.

Inserting \eqref{D} in \eqref{Gamma}, in the case of an initial Fock state 
$\rho =|n\rangle\langle n|$ we get:
\begin{equation}
\label{Gamma-Fock}
\Gamma_{\rm Fock} =-\langle n |\mathcal{D}\big[| n\rangle\langle n |\big]| n\rangle 
=\lambda^2\Big[(n+1)(N-n)\, h(\omega_n)+n(N-n+1)\, \kappa(\omega_{n-1})\Big]\ ,
\end{equation}
while for the coherent like states \eqref{psiN} one finds
\begin{equation}
\label{Gamma-coherent}
\begin{split}
\Gamma_{\rm coherent} & =-\langle \Psi_N|
\mathcal{D}\big[|\Psi_N\rangle \langle \Psi_N|\big]|\Psi_N\rangle \\
& =\lambda^2\sum_{n=0}^N(n+1)(N-n)\Big\{|C_n|^2\left(1-|C_{n+1}|^2\right)\, h(\omega_n)
+ |C_{n+1}|^2\left(1-|C_n|^2\right)\, \kappa(\omega_n)
\Big\}\ .
\end{split}
\end{equation}
Since $N\gg\bar n\gg 1$, following arguments similar to those leading to \eqref{coherent},
the contributing terms to the sum 
can be approximated as $|C_{n+1}|\approx|C_n|\approx 10^{-4}$, so that
$(1-|C_n|)\approx 1$; as a consequence, one can write
\begin{equation}
\label{Gamma-coherent2}
\Gamma_{\rm coherent}\approx \lambda^2\sum_{n=0}^N(n+1)(N-n)|C_n|^2
\big[h(\omega_n)+ \kappa(\omega_n)\big]\ .
\end{equation}

To proceed further, we shall consider a very common instance of environment,
that of a heat bath having two-point correlation functions of exponentially decaying form,
$$
\langle B^\dagger(t)\, B\rangle_E=\langle B(t)\, B^\dagger\rangle_E=g^2\, \exp(-|t|/\tau_{E})\ ,
$$
where $\tau_E$ is the characteristic time scale of the environment and $g^2\sim|\langle B^2\rangle_E|$ is a constant measuring the strength of the bath correlations.
Recalling the expression of the frequency $\omega_n$ in \eqref{omega-n}, 
it is convenient to introduce the new integer variable $k$, by writing $n=\bar{n}+k$;
at resonance, $n_g=1/2$, one then has $\omega_{\bar{n}+k}=2E_C\,k$. 
The Fourier transforms \eqref{Fourier}
can now be explicitly computed, giving
\begin{equation}
h_k \equiv h(\omega_{\bar{n}+k})=\kappa(\omega_{\bar{n}+k})=
\int_{-\infty}^\infty\mathrm{d}t\, g^2\, e^{-|t|/\tau_{E}+i\omega_{\bar n+k}t}=
\frac{2\, g^2\,\tau_E}{1+\left(rk\right)^2}\ ,
\end{equation}
where
\begin{equation}\label{eq:r} 
r\equiv\frac{\tau_E}{\tau_C} =2E_C{\tau_E}\ ,
\end{equation}
with $\tau_C\equiv (2E_C)^{-1}$ being the characteristic time of oscillations due to Coulomb interaction.%
\footnote{In the experiment reported in \cite{Naka1,Naka2},
the charging energy is around $E_C\approx 500\,\mu $eV, so that $\tau_C\sim 10^{-11}$~s.}

With these results,
using the central limit theorem
to approximate $|C_{\bar n +k}|^2$ by a Gaussian distribution,%
\footnote{Substituting \eqref{psiL}, \eqref{psiR} in the definition \eqref{Cn}, one finds
$|C_n|^2=\frac{N!}{n!(N-n)!}\left(\frac{\bar{n}}{N}\right)^{n}\left(1-\frac{\bar{n}}{N}\right)^{N-n}$,
since $\langle \hat{n}_L\rangle \approx \bar n$. Using the Stirling formula
for the various factorials, one further gets
$|C_{\bar{n}+k}|^2\approx\frac{1}{\sqrt{2\pi}}\sqrt{\frac{N}{n(N-n)}}\ g(k)$
where, in the limit $N\gg \bar n$, the function $g(k)$ can be very well approximated by a Gaussian 
distribution (for further details, see \cite{John}).}
$\Gamma_{\rm coherent}$ in \eqref{Gamma-coherent2} can be cast in the following form:
\begin{equation}
\label{Gamma-sum}
\Gamma_{\rm coherent}\approx \lambda^2\sum_{k=-\bar{n}}^{N-\bar{n}}(\bar{n}+k+1)(N-\bar{n}-k)\frac{2\, g^2\,\tau_E}{1+\left(rk\right)^2}\frac{\mathrm{e}^{-\frac{k^2}{2\bar{n}}}}{\sqrt{2\pi\, \bar{n}}}\ ,
\end{equation}
that, in turn, due to the large numbers involved, is  well approximated by
\begin{equation}
\Gamma_{\rm coherent}\approx \lambda^2\left(2\, g^2\,\tau_E\right)\bar{n}(N-\bar{n})\ f(\sqrt{\bar n}\, r)\ ,
\qquad
f(z)\equiv \frac{1}{\sqrt{2\pi}}\int_{-\infty}^\infty\mathrm{d}y\, 
\frac{\mathrm{e}^{-y^2/2}}{1+\left(z\,y\right)^2}\ .
\end{equation}
One can similarly evaluate the expression of the decay rate in the case of occupation number states,
obtaining from \eqref{Gamma-Fock}:
\begin{equation}
\label{Fock-1}
\Gamma_{\rm Fock}\approx \lambda^2\left(2\, g^2\,\tau_E\right)\bar{n}(N-\bar{n})\ .
\end{equation}
The integrand in $f(z)$ above is a product of two decaying functions with different characteristic scales, namely a Gaussian with standard deviation equal to one  and a Lorentzian with width $1/z$. Thus, in the large $z$ limit the integral is dominated by the Lorentzian. Hence, for $\sqrt{\bar{n}}r\gg 1$, we get
\begin{equation}
\frac{\Gamma_{\rm Fock}}{\Gamma_{\rm coherent}}=\frac{1}{f(\sqrt{\bar{n}}r)}\propto\sqrt{\bar{n}}r\ ,
\end{equation}
showing that the decay rate in the case of occupation number states results much larger 
than the corresponding one for coherent states. 

However, to be experimentally relevant, this
result needs to be interpreted within the actual conditions of a typical setup;
in particular, one needs to estimate the magnitude of these decay rates and compare them
with the smallest, characteristic energy scale of the device under study, {\it i.e.} the tunneling
energy $E_J$.

Indeed, as already mentioned, in order for the noise effects to be observable, the strength
of the interaction describing the coupling of the SCB with the environment 
should be of the same order of magnitude of the tunneling term. Recalling the form of the
corresponding Hamiltonian pieces \eqref{h0} and \eqref{h}, this condition roughly means
$\lambda g\approx K$,%
\footnote{This estimate comes from imposing
$\langle H_J\rangle\simeq\langle H_I\rangle$, where $H_J$ is the tunneling term in \eqref{h0}
and $H_I$ is the interaction term in \eqref{h}. Indeed,
$\langle H_J\rangle \simeq K\langle \hat{a}_L\hat{a}_R^\dag\rangle$,
while $\langle \lambda H_I\rangle \simeq \lambda \langle B\rangle\langle \hat{a}_L\hat{a}_R^\dag\rangle\sim\lambda g\langle \hat{a}_L\hat{a}_R^\dag\rangle$, since
$g\simeq\sqrt{|\langle B^2\rangle|}$.}
or equivalently $\lambda g\sqrt{\bar n(N-\bar n)}\approx E_J$.
Then, from the expression \eqref{Fock-1}, one immediately gets the estimate:
\begin{equation}
\label{eq:estimate}
\frac{\Gamma_{\rm Fock}}{E_J}\approx \frac{E_J}{E_C}\, r\ .
\end{equation}
The ratio $E_J/E_C$ is fixed in any experimental situation. For instance, in the setup described in 
\cite{Naka1, Naka2}, one has
\begin{align*}
E_J &\approx 50\; \mu\text{eV}\approx 10^{10}\;\text{s}^{-1},\\
E_C &\approx 500\; \mu\text{eV}\approx 10^{11}\;\text{s}^{-1},
\end{align*}
so that $E_J/E_C\approx 1/10$. As a consequence, in the relevant regime $\sqrt{\bar n}\, r\gg 1$, 
the decay rate $\Gamma_{\rm Fock}$ might be close to $E_J$ 
and therefore the effects of the environment visible. 
In particular, by choosing an environment for which $r$ is close to one,%
\footnote{This choice is compatible with the weak coupling limit.
As mentioned before, the application of this procedure requires that the decoherence
time scale of the slow subsystem dynamics $\tau\equiv 1/\Gamma$ be much greater of both
the characteristic decay time $\tau_E$ of correlations in the environment and the intrinsic
time scale of the free system dynamics $\tau_C=1/2E_C$. One checks \cite{John} that with the choice 
$r\approx 1$ both conditions are satisfied.}  
one obtains $\Gamma_{\rm Fock}\approx E_J/10$, or equivalently
a decay time of order $\tau_{\rm Fock}=1/\Gamma_{\rm Fock}\approx 10/E_J\approx 10^{-9}$ s. 

On the other hand, for the coherent states one gets 
\begin{equation}\label{eq:coh-time-estimate}
\frac{\Gamma_{\rm coherent}}{E_J}\approx \frac{1}{\sqrt{\bar n}}\frac{E_J}{E_C}\ ,
\end{equation}
which is independent of $r$ and, in the case of the setup in \cite{Naka1, Naka2}, 
yields a decay time of order
$\tau_{\rm coherent} =1/\Gamma_{\rm coherent}\sim 10^{-5}\ {\rm s}$.
This means that if the charge oscillations
are a manifestation of macroscopic quantum effects, their damping would become evident after just few oscillation periods, while if they are semiclassical in nature, they will persist
for very long times. 
In the specific case of the experiment described in \cite{Naka1, Naka2}, 
by doubling or tripling the observation time
in presence of a suitably engineered environment, damping of the charge oscillations 
must be observed if the oscillations are indeed a manifestation of quantum coherence.

As a final remark, note that the ratio $\omega_c/\omega_q$ between the charge oscillation frequencies
in \eqref{c-frequency} and \eqref{q-frequency} behaves as
$(E_C/E_J)^{1/2}$; thus, by modifying
the experimental conditions so that to
increase the ratio $E_C/E_J$,
one can eventually distinguish between the two types of explanation 
of the observed charge oscillations through a frequency measure.
Nevertheless, the alternative approach here proposed of adding noise to the device might be, in principle, 
more efficient in discriminating between the two models. 
Indeed, the damping time for Fock states increases linearly with $E_C/E_J$, 
while the ratio of the frequencies above only with the square root.

\section{Conclusions}

Reviewing the phenomenology of charge oscillations in a superconducting
Cooper pair box, we have seen that they might be modelled in two different ways, 
either as a manifestation of macroscopic
quantum coherence (leading to the widely accepted qubit interpretation) 
or as the result of a semiclassical mean-field approach (through a Gross-Pitaevskii like equation).
However, the response of the SCB to external noise, generated by a weakly coupled environment,
is found to be very different in the two models. Both approaches predict damping of the oscillations,
but the decay rate in the case of qubit (Fock) states differs by a factor $\sqrt{\bar n}\, 
r$ from that of the mean-field (coherent like) ones. 
In the physically relevant regime $\sqrt{\bar n}\, r\gg 1$, one then expects the Fock states to decay much faster than the coherent ones. 

This result might provide a way to distinguish experimentally between the two possible interpretations
of the observed SCB charge oscillations. The idea is to couple an SCB with an externally controlled
environment, satisfying the conditions of the weak coupling limit approximation. As soon as the interaction
with the environment is switched on and therefore noise is injected into the nanodevice, 
damping of the charge oscillations should become visible if these are a manifestation of
macroscopic quantum coherence, as described by the qubit model; on the other hand, 
if they survive for long times, this would be an indication
of their semiclassical origin, as described by the mean-field approach.

One can easily evaluate the visibility of the damping effects in the specific setup
described in \cite{Naka1,Naka2}. For the physically relevant case of a heat bath with $r\approx 1$,
we have found that the decay time for occupation number, 
Fock states can be estimated in about $10^{-9}\ {\rm s}$. 
As a consequence, if due to macroscopic quantum coherence, the decoherence effects must be visible after about
a few oscillation periods. In the case of coherent states, we have instead obtained
a much longer decay time of about $10^{-5}\ {\rm s}$. Therefore, the possibility to experimentally
distinguish between the two types of interpretation with already existing setups appears quite realistic.

\section*{Acknowledgments}

JR-G gratefully acknowledges financial support from COLCIENCIAS under  
Research Grant No. 1106-14-17903 and the Mazda Foundation for the Arts and Sciences, and also thanks John Henry Reina for encouragement and support.


\begin{thebibliography}{99}

\bibitem{Naka1}
Y. Nakamura, Yu. Pashin and J.S. Tsai, Nature \textbf{398} (1999) 786

\bibitem{Naka2}
Y. Nakamura, Yu.A. Pashkin and J.S. Tsai, Phys. Rev. Lett. \textbf{87}
(2001) 246601

\bibitem{Yama}
T. Yamamoto, Y. A. Pashkin, O. Astafiev and Y. Nakamura and J.S. Tsai, 
Nature {\bf 425} (2003) 941

\bibitem{Berkeley}
A. J. Berkeley, H. Xu, R. C. Ramos, M. A. Gubrud, F. W. Strauch, P. R. Johnson, 
J. R Anderson, A. J. Dragt, C. J. Lobb and F. C. Wellstood,
Science {\bf 300} (2003) 1548

\bibitem{Steffen}
M. Steffen, M. Ansmann, R. C. Bialczak, N. Katz, E. Lucero, R. McDermott, M. Neeley, 
E. M. Weig, A. N. Cleland and J. M. Martinis, Science {\bf 313} (2006) 1423 

\bibitem{Leggett}
A. J. Leggett, in {\it Chance and Matter}, edited by J. Souletie, J. Vannimenus, and R. Stora. 
(Elsevier, Amsterdam, 1987)

\bibitem{Grabert}
{\it Single Charge Tunneling: Coulomb Blockade Phenomena in Nanostructures},
H. Grabert and M. H. Devoret, eds., NATO Science Series: B, (Plenum Press, New York, 1992). 

\bibitem{Makhlin}
Y. Makhlin, G. Sch\"on, and A. Schnirman, Rev. Mod. Phys. {\bf 73} (2001) 357

\bibitem{Wendin}
G. Wendin, V.S. Schumeiko, Superconducting Quantum Circuits, Qubits and Computing,
cond-mat/0508729

\bibitem{Alicki1}
R. Alicki, Quantumness of Josephson Junctions Reexamined, {\tt quant-ph/0610008}

\bibitem{Alicki2}
R. Alicki, Open Sys. and Inform. Dyn. {\bf 14} (2007) 223

\bibitem{Gronbech1}
N. Gronbech-Jensen, M. G. Castellano, F. Chiarello, M. Cirillo, C. Cosmelli, 
L. V. Filippenko, R. Russo and G. Torrioli, 
Phys. Rev. Lett. {\bf 93} (2004) 107002

\bibitem{Gronbech2} N. Gronbech-Jensen and M. Cirillo,
Phys. Rev. Lett. {\bf 95} (2005) 067001

\bibitem{Gronbech3}
J.E. Marchese, M. Cirillo and N. Gronbech-Jensen,
Open. Syst. Inform. Dynam. {\bf 14} (2007) 189

\bibitem{K}
V. Gorini, A. Frigerio, M. Verri, A. Kossakowski and E.G.C. Sudarshan,
Rep. Math. Phys. {\bf 13} (1978) 149

\bibitem{AL}
R. Alicki and K. Lendi, {\it Quantum Dynamical Semigroups and 
Applications}, Lect. Notes Phys. {\bf 286}, (Springer-Verlag, Berlin, 1987)

\bibitem{BF}
F. Benatti and R. Floreanini, Int. J. Mod. Phys. B \textbf{19}
(2005) 3063

\bibitem{BP}
H.-P. Breuer and F. Petruccione, {\it The Theory of Open
Quantum Systems} (Oxford University Press, Oxford, 2002)

\bibitem{ABF}
R. Alicki, F. Benatti, and R. Floreanini, Charge oscillations in superconducting
nanodevices coupled to external environments,
Phys. Lett. A, to appear, {\tt arXiv:0711.0812}

\bibitem{Gorini}
V. Gorini, A. Kossakowski, J. Math. Phys. \textbf{17} (1976) 1298

\bibitem{Davies}
E. B. Davies, Comm. Math. Phys. {\bf 39} (1974) 91; 
Math. Ann. {\bf 219} (1976) 147

\bibitem{Choi}
M.-S. Choi, R. Fazio, J. Siewert and C. Bruder,
Europhys. Lett. {\bf 52} (2001) 251

\bibitem{Cirac}
D. Jaksch, C. Bruder, J.I. Cirac, C.W. Gardiner and P. Zoller,
Phys. Rev. Lett. \textbf{81} (1998) 3108

\bibitem{Lewe}
M. Lewenstein, A. Sanpera, V. Ahufinger, B. Damski, A. Sen De nd U. Sen,
Adv. in Phys. \textbf{56}  (2007) 243

\bibitem{John}
John Realpe-G\'omez, {\it Stability of the Single Cooper Pair Box Interacting 
with an Environment}, Diploma thesis, ICTP, 2007


\end{thebibliography}
\end{document}